\documentclass{article}
\usepackage{spconf,amsmath,graphicx}

\usepackage{floatrow}
\DeclareFloatFont{normalsize}{\normalsize}
\usepackage{amsfonts}
\usepackage{amssymb}
\usepackage{amsthm}   
\usepackage{subfig}
\usepackage{tabularx}
\usepackage{multirow}
\usepackage{makecell}
\usepackage{adjustbox}
\usepackage{url}
\usepackage{empheq}

\usepackage{caption}

\title{ECG HEART-BEAT CLASSIFICATION USING MULTIMODAL IMAGE FUSION}
\name{Zeeshan Ahmad$^1$, Anika Tabassum$^2$, Ling Guan$^3$, Naimul Khan$^4$}
\address{$^1$ $^3$ $^4$ Department of Electrical, Computer and Biomedical Engineering, Ryerson University, Toronto, ON.\\$^2$Master of Data Science program, Ryerson University, Toronto, ON. }
\begin{document}
%
\maketitle
\begin{abstract}
	
In this paper, we present a novel Image Fusion Model (IFM) for ECG heart-beat classification to overcome the weaknesses of existing machine learning techniques that rely either on manual feature extraction or direct utilization of 1D raw ECG signal. At the input of IFM, we first convert the heart-beats of ECG into three different images using Gramian Angular Field (GAF), Recurrence Plot (RP) and Markov Transition Field (MTF) and then fuse these images to create a single imaging modality. We use AlexNet for feature extraction and classification and thus employ end-to-end deep learning. We perform experiments on PhysioNet’s MIT-BIH dataset for five different arrhythmias in accordance with the AAMI EC57
standard and on PTB diagnostics dataset for myocardial infarction (MI) classification. We achieved an state-of-an-art results in terms of prediction accuracy, precision and recall.

\end{abstract}
\section{Introduction} \label{sec:intro}

ECG heart-beat investigation is important for early diagnosis of cardiovascular diseases such as arrhythmias and myocardial infarction (MI) as ECG is the best source to provide electrophysiological pattern of depolarization and repolarization of the heart muscles. Arryhthmias is a heart rhythmic problem which happens when electrical signals coordinating heart-beats causes heart to beat irregularly. Myocardial Infarction, also called heart attack, is a serious threat to human life and is caused due to the blockage
of oxygen-rich blood to the heart, thus resulting in severe cardiac arrest and can be dangerous for patient's life~\cite{acharya2005study}.

\let\thefootnote\relax\footnote{© 2021 IEEE. Personal use of this material is permitted. Permission from IEEE must be obtained for all other uses, in any current or future media, including reprinting/republishing this material for advertising or promotional purposes, creating new collective works, for resale or redistribution to servers or lists, or reuse of any copyrighted component of this work in other works.}

ECG is a reliable tool to interpret the cardiovascular condition. However, ECG heart-beat classification is an uphill task for researchers due to the complex and non-stationary nature of the ECG signal~\cite{zhang2014heartbeat}. Thus, computer based approaches for intellegent and automatic identification of abnormalities in heart-beat are in demand.

Earlier methods for heart-beat classification using ECG signal were dependent upon manual feature extraction using signal processing~\cite{pasolli2010active} and statistical techniques~\cite{bhaskar2015performance}. The disadvantages with these conventional methods are the separation of feature
extraction part and pattern classification part and the expert knowledge about the input data
and selected features~\cite{sidek2014ecg}. Moreover, hand-crafted features may
not invariant to noise, scaling and translations and thus can lead to the problem of generalization on unseen data.

Outstanding performance of deep learning models especially the performance of CNN
in computer vision~\cite{ahmad2019humanactionrec} and image classification~\cite{krizhevsky2012imagenet} has gained attention of researchers since the deep learning models are capable of automatically learning hierarchical features directly from the data and promote end-to-end learning. Recent deep learning models use 1D ECG signal or 2D representation of ECG by transforming ECG signal to images. 2D representation of ECG provides more accurate heart-beat classification compared to 1D~\cite{huang2019ecg}. Furthermore, multimodal fusion of 2D representation of ECG achieved highest accuracy as compared to single ECG modality~\cite{ahmad2020multi}. 

However, existing fusion methods rely  on concatenation or decision level fusion~\cite{de2004automatic}. There is room for improvement in fusion techniques that can provide better results while not sacrificing efficiency. To address the shortcomings of existing fusion models for ECG heartbeat classification, in this paper, we
propose image fusion model (IFM) that fuses three gray scale images to form a single three channel image  containing both static and dynamic features of input images and thus achieved better classification results while taking care of dimensionality as well.

The key contributions of the presented work are:

\begin{enumerate}	
	
	\item We propose a novel image fusion model (IFM) that fuses three gray scale images to form a single three channel image  containing features of input images and thus achieved better classification results. The proposed model not only promotes the end-to-end learning but also computationally efficient by keeping dimensionality of the fused features equal to the dimensionality of single modality feature.

	\item At the input of the IFM, converting heartbeats of ECG signal to images using Gramian Angular Field (GAF), Recurrence Plot (RP) and Markov Transition Field (MTF), preserves the temporal correlation among the samples of time series and thus we achieve better classification performance as compared to the existing methods of transforming ECG to images using spectrograms or methods involving time-frequency analysis.

\end{enumerate}	

\section{Related Work} \label{sec:related work}

Since 2D form of ECG signals such as images perform better than 1D raw ECG data~\cite{huang2019ecg}, a big chunk of existing work related to ECG heart-beat classification includes conversion of ECG to images. In~\cite{hao2019spectro}, ECG signal is converted into spectro-temporal images. Multiple dense CNNs were used to capture both beat-to-beat and single-beat information for analysis. Authors in~\cite{oliveira2019novel} converted  heart-beats of ECG signals to images using wavelet transform. A six layer CNN was trained on these images for heartbeat classification. In~\cite{diker2019novel}, well known pretrained CNNs
such as AlexNet, VGG-16 and ResNet-18 are trained on spectrograms obtained from ECG. Using a transfer learning approach, the highest accuracy of 83.82\% is achieved by AlexNet. In~\cite{izci2019cardiac}, ECG heart-beats were transformed to 2D grayscale images and then CNN was used for feature extraction and classification.

 Multimodal fusion enhances the performance as compared to individual modalities by integrating
complementary information from the modalities. A Deep Multi-scale Fusion CNN (DMSFNet) is proposed in~\cite{fan2018multiscaled} for arrhythmia detection. Proposed model
consists of backbone network and two different scale-specific networks. Features obtained from two scale specific networks are fused using a spatial attention module. CNN and attention module based multi-level feature fusion framework is proposed in~\cite{wang2019multi} for multiclass arrhythmia detection. Heart-beat classification is performed by extracting features from various layers of CNN. It is observed that combining
the attention module and CNN improves the classification results.

The shortcoming in the existing fusion methods is that
they depend mostly on concatenation fusion. Concatenation
creates the problem of the curse of dimensionality and high
computational cost, results in the degradation of accuracy.
To address the shortcomings of existing works, in this paper,
we propose a novel image fusion model that fuses input images to form a single three channel image  containing both static and dynamic features of input images and thus achieved better classification results while taking care of dimensionality as well.

\begin{figure}
	\centering
	\includegraphics[width=\linewidth]{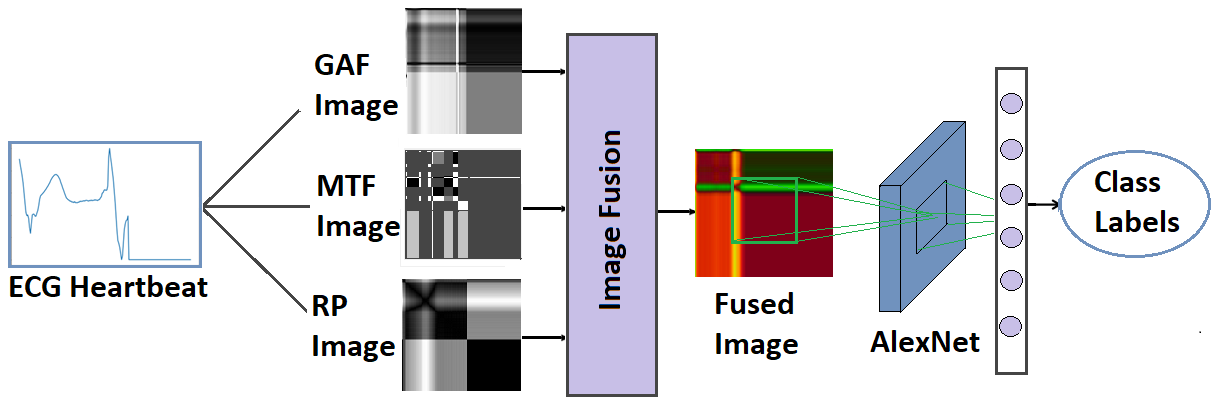}
	\caption{Complete Overview of the proposed Image Fusion Model.} 
	\label{fig: MIF framework}
\end{figure}

\section{Proposed Method}\label{sec:proposed method}

In this section we will explain the proposed image fusion model (IFM) as shown in Fig.~\ref{fig: MIF framework}.

\subsection{ECG Signal to Image Transformation} \label{ sec:ECG to image formation}

At the input of the proposed model, we transform the heartbeats of input ECG signal into GAF, RP and MTF images. 

\subsubsection{Image formation by Gramian Angular Field (GAF)}\label{sec:gaf image formation}

The image formed by Gramian Angular Field (GAF) represents time series in a polar coordinate system rather than conventional cartesian coordinate system.

Let $X\in\mathbb{R}$ be a real valued time series of $n$ samples such that $X = \{x_1, x_2, x_3,...,x_i,...,x_n\}$. We rescaled $X$ to $X_s$ so that the value of each sample in $X_s$ falls between 0 and 1.
Now we represent the rescaled time series in polar coordinate system by encoding the value as the angular cosine and the time stamp as the radius. This encoding can be understood by the following equation.

\begin{empheq}[right=\empheqrbrace]{equation}
\begin{split}
\phi = arccos(x_{i0}) \\
r = \frac{t_i}{N}
\end{split}
\end{empheq}
where $x_{i0}$ is the rescaled $ith$ sample of the time series, $t_i$ is the time stamp and $N$ is a constant factor to regularize the span of the polar coordinate system~\cite{wang2015imaging}.
The angular perspective of the encoded image can be fully utilized by cosidering the sum/difference between each point to identify the temporal correlation within different time intervals. In this paper we used a  summation method for Grammian Angular Field and is explained by the following equation

\begin{equation}
GAF = cos(\phi_i + \phi_k)
\end{equation}

\begin{figure}[h]
	\centering
	\includegraphics[width=0.9\linewidth]{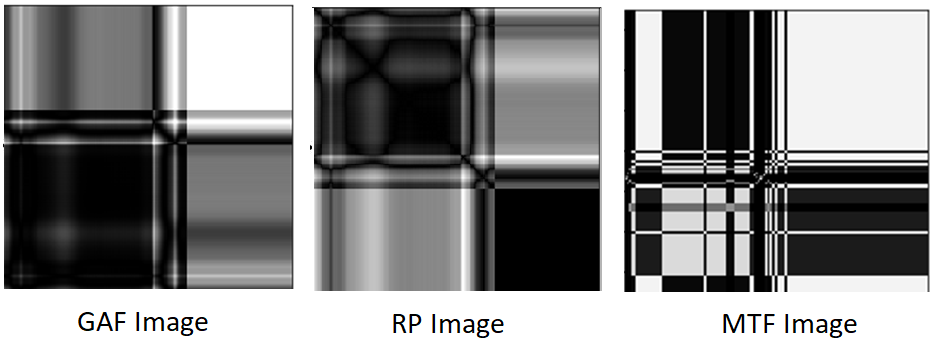}
	\caption{GAF, RP and MTF Images.}
	\label{fig: GAF Image}
\end{figure}

The image formed by GAF for a single heartbeat is shown in Fig~\ref{fig: GAF Image}.

\subsubsection{Image formation by Recurrence Plot (RP)}\label{sec:rp image formation}

Periodicity and irregular cyclicity are the key recurrent behaviors of time series data. The recurrence plots are used as visualization tool for observing the recurrence structure of a time-series~\cite{eckmann1995recurrence}. A recurrence plot (RP) is an image obtained from a multivariate time-series, representing the distances between each time point.

Let $q(t)\in\mathbb{R}^{d}$ be a multi-variate time-series. Its recurrence plot is defined as

\begin{equation}\label{ eq:RP}
RP = \theta(\epsilon - ||q(i) - q(j)||)
\end{equation}
In equation~\ref{ eq:RP}, $\epsilon$ is threshold and $\theta$ is called heaviside function.

Since our ECG signal is univariate, for our case, $d=1$.
The image formed by RP is shown in Fig~\ref{fig: GAF Image}.

\subsubsection{Image formation by Markov Transition Field (MTF)} \label{sec: mtf image formation}

We used the method described in~\cite{wang2015imaging} to encode ECG signal into images. Consider the time series $X\in\mathbb{R}$ such that $X = \{x_1, x_2, x_3,...,x_l,...,x_n\}$. The first step is to identify its $Q$ quantile bins and
assign each $x_l$ to the corresponding bins $q_k (k\epsilon[1, Q])$.
Next step is to construct a $Q \times Q$  weighted adjacency matrix $W$
by counting transitions among quantile bins in the manner of
a first-order Markov chain along the time axis. The Markov transition field matrix is given by 

\begin{equation}\label{eq:Markov transition field matrix}
M=
\begin{bmatrix}
w_{lk|x_1\epsilon q_l,x_1\epsilon q_k}  & \dots & w_{lk|x_1\epsilon q_l,x_n\epsilon q_k}\\
w_{lk|x_2\epsilon q_l,x_1\epsilon q_k} & \dots  & w_{lk|x_2\epsilon q_l,x_n\epsilon q_k}\\
\vdots      &     \ddots &          \vdots    \\
w_{lk|x_n\epsilon q_l,x_1\epsilon q_k}  & \dots & w_{lk|x_n\epsilon q_l,x_n\epsilon q_k}

\end{bmatrix}
\end{equation}
where $w_{lk}$ is the frequency with which a point in quantile $q_k$ is followed by a point in quantile $q_l$. 

We use 10 bins for the discretization and encoding of ECG hearbeats into images. The image formed by MTF is shown in Fig~\ref{fig: GAF Image}.

\subsection{Image Fusion Model}

 After image formation from ECG heart-beat, we combine these three gray scale images to form a triple channel image (GAF-RP-MTF) which contains both static and dynamic features of the input images and thus enhance classification performance. A triple channel image is a colored image in which GAF, RP and MTF images are considered as three orthogonal channels like three different colors in RGB image space. We use AlexNet, (CNN based model)~\cite{krizhevsky2012imagenet} for feature extraction and classification tasks and thus employ end-to-end deep learning where feature extraction and classification parts are embedded in a single network as shown in Fig.~\ref{fig: MIF framework}.

\begin{table}
	\centering
	\scalebox{0.9}{
		\begin{tabular}{c c c c }	
			\hline\hline
			\textbf{Modalities} & \textbf{Accuracies\%} & \textbf{Precision\%} & \textbf{Recall\%}  \\\hline\hline
			GAF Images only  &  97.3 & 85 & 91 \\\hline
			RP Images only & 97.2 &  82 &  93  \\\hline
			MTF Images only & 91.5 & 86 &  89  \\\hline	
			Proposed IFM  &  98.6 & 93 & 92 \\\hline

	\end{tabular}}
	\caption{Ablation Study on MIT-BIH Dataset}
	\label{tab : experiments on MIT-BIH Dataset using AlexNet}
\end{table}

\begin{table}[h]
	\centering
	\scalebox{0.9}{
		\begin{tabular}{c c c c }	
			\hline\hline
			\textbf{Modalities} & \textbf{Accuracies\%} & \textbf{Precision\%} & \textbf{Recall\%}  \\\hline\hline
			GAF Images only  &  98.4 & 98 & 96  \\\hline
			RP Images only & 98 &  98 & 94  \\\hline
			MTF Images only & 95.3 & 94 & 89 \\\hline	
			Proposed IFM &  98.4 & 98 & 94  \\\hline

	\end{tabular}}
	\caption{Ablation Study on PTB Dataset.}
	\label{tab : experiments on PTB Dataset using AlexNet}
\end{table}

\begin{table}[h]
	\centering
	\scalebox{0.9}{
		\begin{tabular}{c c c c}	
			\hline\hline
			\textbf{Previous Methods} & \textbf{Accuracies\%} & \textbf{Precision\%} & \textbf{Recall\%} \\\hline\hline
			
			Izci et al.~\cite{izci2019cardiac} & 97.96 & - & -  \\\hline
			Zhao et al.~\cite{zhao2017electrocardiograph} & 98.25 & - & - \\\hline
			Oliveria et al.~\cite{oliveira2019novel} & 95.3 & - & - \\\hline	
			Shaker et al.~\cite{shaker2020generalization} & 98 & 90 & 97.7 \\\hline
			Kachuee et al.~\cite{kachuee2018ecg} & 93.4 & - & - \\\hline
			\textbf{Proposed IFM} & \textbf{98.6} & \textbf{93} & \textbf{92} \\\hline	
				
	\end{tabular}}
	\caption{Comparison of heart beat Classification results of MITBIH Dataset with Previous Methods}
	\label{tab : Comparison of MIT-BIH}
\end{table} 

\begin{table}[h]
	\centering
	\scalebox{0.9}{
		\begin{tabular}{c c c c}	
			\hline\hline
			\textbf{Previous Methods} & \textbf{Accuracies\%} & \textbf{Precision\%} & \textbf{Recall\%} \\\hline\hline
			
			Dicker et al.~\cite{diker2019novel} & 83.82 & 82 & 95  \\\hline
			Kojuri et al.~\cite{kojuri2015prediction} & 95.6 & 97.9 & 93.3  \\\hline
			Kachuee et al.~\cite{kachuee2018ecg} & 95.9 & 95.2 & 95.1 \\\hline
			Liu et al.~\cite{liu2017real} & 96 & 97.37 & 95.4 \\\hline
			\textbf{Proposed IFM} & \textbf{98.4} & \textbf{98} & \textbf{94} \\\hline	
				
	\end{tabular}}
	\caption{Comparison of MI Classification results of PTB Dataset with Previous Methods}
	\label{tab : Comparison of PTB}
\end{table}

\section{Experiments and Results} \label{sec:Experiment and results}

We experiment on PhysioNet MIT-BIH Arrhythmia dataset~\cite{moody2001impact} for heartbeat classification and PTB Diagnostic ECG dataset~\cite{bousseljot1995nutzung} for MI classification. For our experiments, we used ECG lead-II re-sampled to the sampling frequency of 125Hz as the input.

We resize the images to 227 x 227 to perform experiments with AlexNet. A drop out ratio of 0.5, momentum of 0.9 and $L_2$ regularization of 0.004 was used. we trained AlexNet till the validation loss stops decreasing further. The experiemntal results are discussed in section~\ref{sec:discussion}.

\subsection{PhysioNet MIT-BIH Arrhythmia Dataset} \label{sec : physionet data}

Forty seven subjects were involved during the collection of ECG signals for the dataset. The data was collected at the sampling rate of 360Hz and each beat is annotated by at least two experts. Using these annotations, five different beat categories are created in accordance with Association for the Advancement
of Medical Instrumentation (AAMI) EC57 standard~\cite{association1998testing}.

The original dataset has 21892 heartbeats, each of which is a 187-point time series. Since there is a class-imbalanced in the  dataset as apparent from the numbers, we applied SMOTE~\cite{chawla2002smote} to upsample the minority classes. The final training and testing samples are 152456 and 21890 respectively.

 The results of experiments 
in terms of recognition accuracies and their comparison with previous state-of-the-art are shown in Tables~\ref{tab : experiments on MIT-BIH Dataset using AlexNet} and~\ref{tab : Comparison of MIT-BIH}.

\subsection{PTB Diagnostic ECG dataset}

Two hundred and ninty (290) subjects took part during collection of ECG records for PTB Diagnostics dataset. 148 of them are diagnosed as MI, 52 healthy control, and the rest are diagnosed with 7 different diseases. Each record contains ECG signals from 12 leads sampled at the frequency of 1000Hz. However, in this paper, we used ECG lead II, and worked with healthy control and MI categories. 

 For experiments with AlexNet, the training and testing samples are 21892 and 2911 respectively. The results of experiments
in terms of recognition accuracies and their comparison with previous state of art are shown in Tables~\ref{tab : experiments on PTB Dataset using AlexNet} and~\ref{tab : Comparison of PTB}.

\section{Discussion}\label{sec:discussion}

The ablation study on both datasets prove that fused three channel image achieved higher accuracy than using single image modality as shown in Tables~\ref{tab : experiments on MIT-BIH Dataset using AlexNet} and~\ref{tab : experiments on PTB Dataset using AlexNet}. Furthermore, Tables~\ref{tab : Comparison of MIT-BIH} and~\ref{tab : Comparison of PTB} show that the proposed fusion method achieved state-of-the-art results and beat previous methods, that depends upon concatenation fusion, in terms of recognition accuracy, precision and recall.

 MTF requires the data to be discretized into $Q$ quantile bins to calculate the $Q \times Q$ Markov transition matrix, therefore the size of MTF images is $Q \times Q$. In this paper $Q = 10$. Thus, the size of MTF images is 10 x 10. For data fusion and to train the AlexNet, we resize 10 x 10 image to 227 x 227. This resizing causes redundancy in MTF images and is likely the reason why there is a drop in recall 
shown in Tables~\ref{tab : Comparison of MIT-BIH} and~\ref{tab : Comparison of PTB}.

\section{Acknowledgement}
Financial support from NSERC and Dapasoft Inc. (CRDPJ529677-18) to conduct the research is highly appreciated.
\section{Conclusion}

In this paper, we propose a novel image fusion model for ECG heart beat classification. At the input of these frameworks, we convert the raw ECG data into three types of images using Gramian Angular Field (GAF), Recurrence Plot (RP) and Markov Transition Field (MTF). We first perform image fusion by combining three input images to create a three channel single image which serve as input to the AlexNet. Experimental results on two publicly available datasets prove the superiority of the proposed method over previous methods.

\bibliographystyle{IEEEbib}

\begin{thebibliography}{10}
	
	\bibitem{acharya2005study}
	U~Rajendra Acharya, N~Kannathal, Lee~Mei Hua, and Leong~Mei Yi,
	\newblock ``Study of heart rate variability signals at sitting and lying
	postures,''
	\newblock {\em Journal of bodywork and Movement Therapies}, vol. 9, no. 2, pp.
	134--141, 2005.
	
	\bibitem{zhang2014heartbeat}
	Zhancheng Zhang, Jun Dong, Xiaoqing Luo, Kup-Sze Choi, and Xiaojun Wu,
	\newblock ``Heartbeat classification using disease-specific feature
	selection,''
	\newblock {\em Computers in biology and medicine}, vol. 46, pp. 79--89, 2014.
	
	\bibitem{pasolli2010active}
	Edoardo Pasolli and Farid Melgani,
	\newblock ``Active learning methods for electrocardiographic signal
	classification,''
	\newblock {\em IEEE Transactions on Information Technology in Biomedicine},
	vol. 14, no. 6, pp. 1405--1416, 2010.
	
	\bibitem{bhaskar2015performance}
	Nitin~Aji Bhaskar,
	\newblock ``Performance analysis of support vector machine and neural networks
	in detection of myocardial infarction,''
	\newblock {\em Procedia Computer Science}, vol. 46, no. 4, pp. 20--30, 2015.
	
	\bibitem{sidek2014ecg}
	Khairul~A Sidek, Ibrahim Khalil, and Herbert~F Jelinek,
	\newblock ``Ecg biometric with abnormal cardiac conditions in remote monitoring
	system,''
	\newblock {\em IEEE Transactions on systems, man, and cybernetics: systems},
	vol. 44, no. 11, pp. 1498--1509, 2014.
	
	\bibitem{ahmad2019humanactionrec}
	Zeeshan Ahmad, Kandasamy Illanko, Naimul Khan, and Dimitri Androutsos,
	\newblock ``Human action recognition using convolutional neural network and
	depth sensor data,''
	\newblock in {\em Proceedings of the 2019 International Conference on
		Information Technology and Computer Communications}, 2019, pp. 1--5.
	
	\bibitem{krizhevsky2012imagenet}
	Alex Krizhevsky, Ilya Sutskever, and Geoffrey~E Hinton,
	\newblock ``Imagenet classification with deep convolutional neural networks,''
	\newblock in {\em Advances in neural information processing systems}, 2012, pp.
	1097--1105.
	
	\bibitem{huang2019ecg}
	Jingshan Huang, Binqiang Chen, Bin Yao, and Wangpeng He,
	\newblock ``Ecg arrhythmia classification using stft-based spectrogram and
	convolutional neural network,''
	\newblock {\em IEEE Access}, vol. 7, pp. 92871--92880, 2019.
	
	\bibitem{ahmad2020multi}
	Zeeshan Ahmad and Naimul~Mefraz Khan,
	\newblock ``Multi-level stress assessment using multi-domain fusion of ecg
	signal,''
	\newblock in {\em 2020 42nd Annual International Conference of the IEEE
		Engineering in Medicine \& Biology Society (EMBC)}. IEEE, 2020, pp.
	4518--4521.
	
	\bibitem{de2004automatic}
	Philip De~Chazal, Maria O'Dwyer, and Richard~B Reilly,
	\newblock ``Automatic classification of heartbeats using ecg morphology and
	heartbeat interval features,''
	\newblock {\em IEEE transactions on biomedical engineering}, vol. 51, no. 7,
	pp. 1196--1206, 2004.
	
	\bibitem{hao2019spectro}
	Chen Hao, Sandi Wibowo, Maulik Majmudar, and Kuldeep~Singh Rajput,
	\newblock ``Spectro-temporal feature based multi-channel convolutional neural
	network for ecg beat classification,''
	\newblock in {\em 2019 41st Annual International Conference of the IEEE
		Engineering in Medicine and Biology Society (EMBC)}. IEEE, 2019, pp.
	5642--5645.
	
	\bibitem{oliveira2019novel}
	Alexandre~Tomazati Oliveira, Euripedes~GO Nobrega, et~al.,
	\newblock ``A novel arrhythmia classification method based on convolutional
	neural networks interpretation of electrocardiogram images,''
	\newblock in {\em IEEE International conference on industrial technology}.
	Piscataway, NJ, 2019.
	
	\bibitem{diker2019novel}
	Aykut Diker, Zafer C{\"o}mert, Engin Avc{\i}, Mesut To{\u{g}}a{\c{c}}ar, and
	Burhan Ergen,
	\newblock ``A novel application based on spectrogram and convolutional neural
	network for ecg classification,''
	\newblock in {\em 2019 1st International Informatics and Software Engineering
		Conference (UBMYK)}. IEEE, 2019, pp. 1--6.
	
	\bibitem{izci2019cardiac}
	Elif Izci, Mehmet~Akif Ozdemir, Murside Degirmenci, and Aydin Akan,
	\newblock ``Cardiac arrhythmia detection from 2d ecg images by using deep
	learning technique,''
	\newblock in {\em 2019 Medical Technologies Congress (TIPTEKNO)}. IEEE, 2019,
	pp. 1--4.
	
	\bibitem{fan2018multiscaled}
	Xiaomao Fan, Qihang Yao, Yunpeng Cai, Fen Miao, Fangmin Sun, and Ye~Li,
	\newblock ``Multiscaled fusion of deep convolutional neural networks for
	screening atrial fibrillation from single lead short ecg recordings,''
	\newblock {\em IEEE journal of biomedical and health informatics}, vol. 22, no.
	6, pp. 1744--1753, 2018.
	
	\bibitem{wang2019multi}
	Ruxin Wang, Qihang Yao, Xiaomao Fan, and Ye~Li,
	\newblock ``Multi-class arrhythmia detection based on neural network with
	multi-stage features fusion,''
	\newblock in {\em 2019 IEEE International Conference on Systems, Man and
		Cybernetics (SMC)}. IEEE, 2019, pp. 4082--4087.
	
	\bibitem{wang2015imaging}
	Zhiguang Wang and Tim Oates,
	\newblock ``Imaging time-series to improve classification and imputation,''
	\newblock in {\em Twenty-Fourth International Joint Conference on Artificial
		Intelligence}, 2015.
	
	\bibitem{eckmann1995recurrence}
	JP~Eckmann, S~Oliffson Kamphorst, D~Ruelle, et~al.,
	\newblock ``Recurrence plots of dynamical systems,''
	\newblock {\em World Scientific Series on Nonlinear Science Series A}, vol. 16,
	pp. 441--446, 1995.
	
	\bibitem{association1998testing}
	Association for the Advancement~of Medical~Instrumentation et~al.,
	\newblock ``Testing and reporting performance results of cardiac rhythm and st
	segment measurement algorithms,''
	\newblock {\em ANSI/AAMI EC38}, vol. 1998, 1998.
	
	\bibitem{zhao2017electrocardiograph}
	Yong Zhao, Xueting Yin, and Yannan Xu,
	\newblock ``Electrocardiograph (ecg) recognition based on graphical fusion with
	geometric algebra,''
	\newblock in {\em 2017 4th International Conference on Information Science and
		Control Engineering (ICISCE)}. IEEE, 2017, pp. 1482--1486.
	
	\bibitem{shaker2020generalization}
	Abdelrahman~M Shaker, Manal Tantawi, Howida~A Shedeed, and Mohamed~F Tolba,
	\newblock ``Generalization of convolutional neural networks for ecg
	classification using generative adversarial networks,''
	\newblock {\em IEEE Access}, vol. 8, pp. 35592--35605, 2020.
	
	\bibitem{kachuee2018ecg}
	Mohammad Kachuee, Shayan Fazeli, and Majid Sarrafzadeh,
	\newblock ``Ecg heartbeat classification: A deep transferable representation,''
	\newblock in {\em 2018 IEEE International Conference on Healthcare Informatics
		(ICHI)}. IEEE, 2018, pp. 443--444.
	
	\bibitem{kojuri2015prediction}
	Javad Kojuri, Reza Boostani, Pooyan Dehghani, Farzad Nowroozipour, and Nasrin
	Saki,
	\newblock ``Prediction of acute myocardial infarction with artificial neural
	networks in patients with nondiagnostic electrocardiogram,''
	\newblock {\em Journal of Cardiovascular Disease Research}, vol. 6, no. 2,
	2015.
	
	\bibitem{liu2017real}
	Wenhan Liu, Mengxin Zhang, Yidan Zhang, Yuan Liao, Qijun Huang, Sheng Chang,
	Hao Wang, and Jin He,
	\newblock ``Real-time multilead convolutional neural network for myocardial
	infarction detection,''
	\newblock {\em IEEE journal of biomedical and health informatics}, vol. 22, no.
	5, pp. 1434--1444, 2017.
	
	\bibitem{moody2001impact}
	George~B Moody and Roger~G Mark,
	\newblock ``The impact of the mit-bih arrhythmia database,''
	\newblock {\em IEEE Engineering in Medicine and Biology Magazine}, vol. 20, no.
	3, pp. 45--50, 2001.
	
	\bibitem{bousseljot1995nutzung}
	R~Bousseljot, D~Kreiseler, and A~Schnabel,
	\newblock ``Nutzung der ekg-signaldatenbank cardiodat der ptb {\"u}ber das
	internet,''
	\newblock {\em Biomedizinische Technik/Biomedical Engineering}, vol. 40, no.
	s1, pp. 317--318, 1995.
	
	\bibitem{chawla2002smote}
	Nitesh~V Chawla, Kevin~W Bowyer, Lawrence~O Hall, and W~Philip Kegelmeyer,
	\newblock ``Smote: synthetic minority over-sampling technique,''
	\newblock {\em Journal of artificial intelligence research}, vol. 16, pp.
	321--357, 2002.
	
\end{thebibliography}

\end{document}